\documentclass[12pt]{article}
\usepackage{graphicx}
\begin{document} 

\bigskip
\centerline
{\bf Meaning and Form in a Language Computer Simulation}

\bigskip
\noindent
S{\o}ren Wichmann$^{1,2}$, Dietrich Stauffer$^3$, Christian Schulze$^3$, 
F. Welington S. Lima $^4$, and Eric W. Holman$^{5,6}$.

\bigskip

\noindent
$^1$ Department of Linguistics, Max Planck Institute for Evolutionary 
Anthropology, Deutscher Platz 6, D-04103 Leipzig, Germany; wichmann@eva.mpg.de

\medskip
\noindent
$^2$ Faculty of Archaeology, Leiden University, P.O. Box 9515, 
2300 RA  Leiden, The Netherlands

\medskip
\noindent
$^3$ Institute for Theoretical Physics, Cologne University, 
D-50923 K\"oln, Euroland

\medskip
\noindent
$^4$ Departamento de F\'{\i}sica,
Universidade Federal do Piau\'{\i}, 54049-550 Teresina - PI, Brazil

\medskip
\noindent
$^5$ Department of Psychology, University of California, Los Angeles, California
90095-1563, USA.

\medskip
\noindent
$^6$ Authors are listed in inverse alphabetical order.
\bigskip

{\small 
Abstract: Thousands of different forms (words) are associated with thousands of 
different meanings (concepts) in a language computer model. Reasonable agreement
with reality is found for the number of languages in a family and the Hamming
distances between languages. 
}

\section{Introduction}

The competition between languages of adult people \cite{compass} has been 
intensively simulated on computers \cite{nettle,abrams} or mathematically
\cite{zanette} for several years. When language structures were
studied, they usually consisted of about a dozen features, often binary 
\cite{kosmidis,schwammle,schulze}; see \cite{ssw} for a review. This
number corresponds roughly to the 47 statistically independent language 
features \cite{holman} in the \textit{World Atlas of Language Structures} 
\cite{wals}, which relate to phonology, morphology, and syntax. 
In contrast, thousands of words are needed in everyday life for thousands of 
different concepts, not counting special terms e.g. from the sciences. 

While the origin of words has already been simulated \cite{baron}, we want to 
simulate the subsequent proliferation and competition between thousands of 
languages, each 
containing thousands of forms for thousands of meanings. In addition we try to 
get realistic statistics for the number of language families containing a 
given number of languages, and for the similarity of languages within one
family and between different families.

\begin{figure}[hbt]
\begin{center}
\includegraphics[angle=-90,scale=0.5]{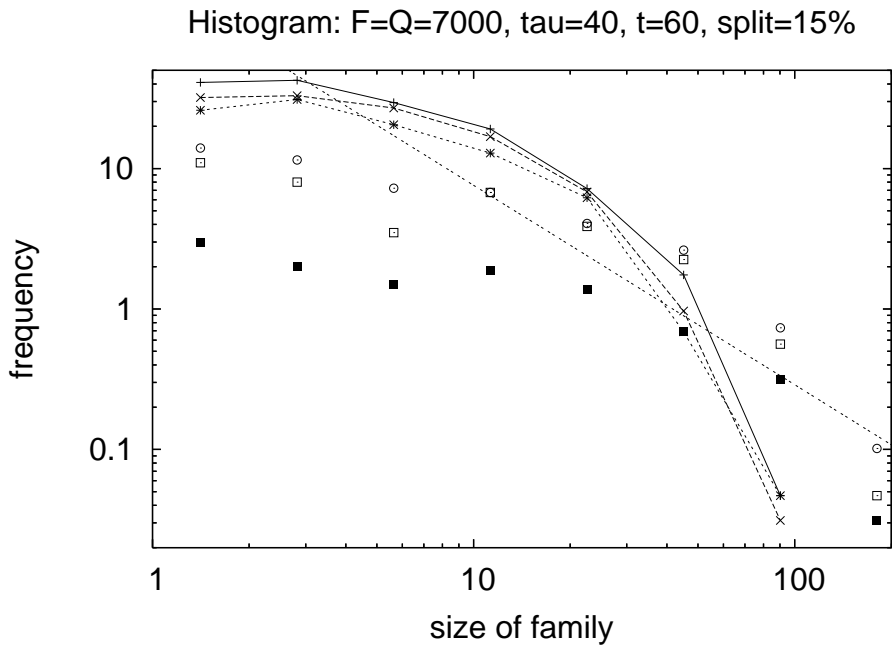}
\end{center}
\caption{Family size distributions. The symbols connected with lines 
correspond to the parameter settings in the headline, while those not connected
with lines have $F=Q=2000$ and $t=100$. In both cases three samples are
shown differing only in the random numbers. The slope of the straight line 
corresponds to the
empirical power law of \cite{wichmann}; see also \cite{zanetteold}.
}
\end{figure}

\begin{figure}[hbt]
\begin{center}
\includegraphics[angle=-90,scale=0.5]{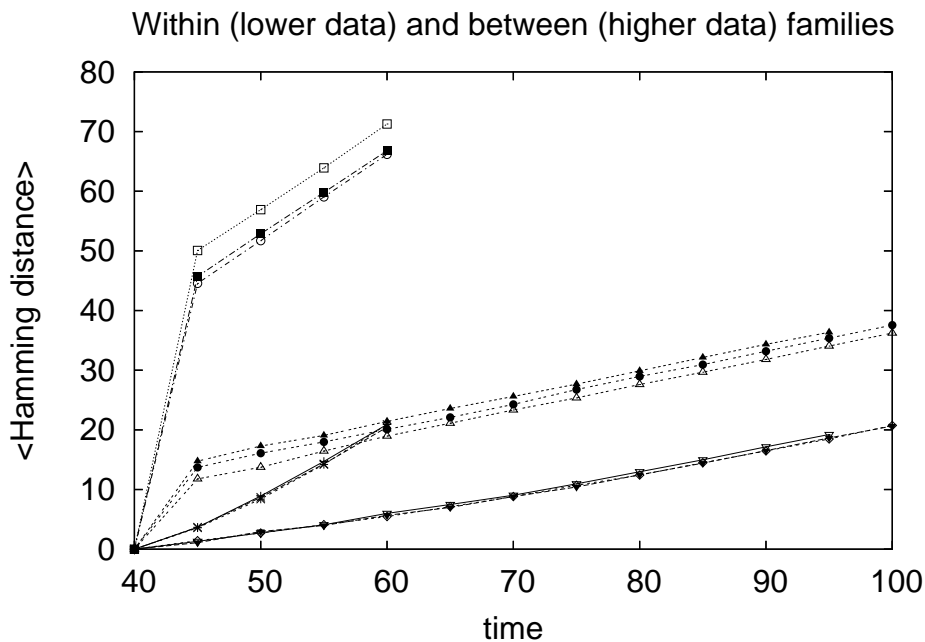}
\end{center}
\caption{Evolution of Manhattan Hamming distances for the simulations of Fig.1. The
simulations up to 60 iterations refer to $F=Q=7000$, those for $t \le 100$
to $F=Q=2000$.  See Fig.14 in \cite{tversky} for similar results.
}
\end{figure}

\begin{figure}[hbt]
\begin{center}
\includegraphics[angle=-90,scale=0.3]{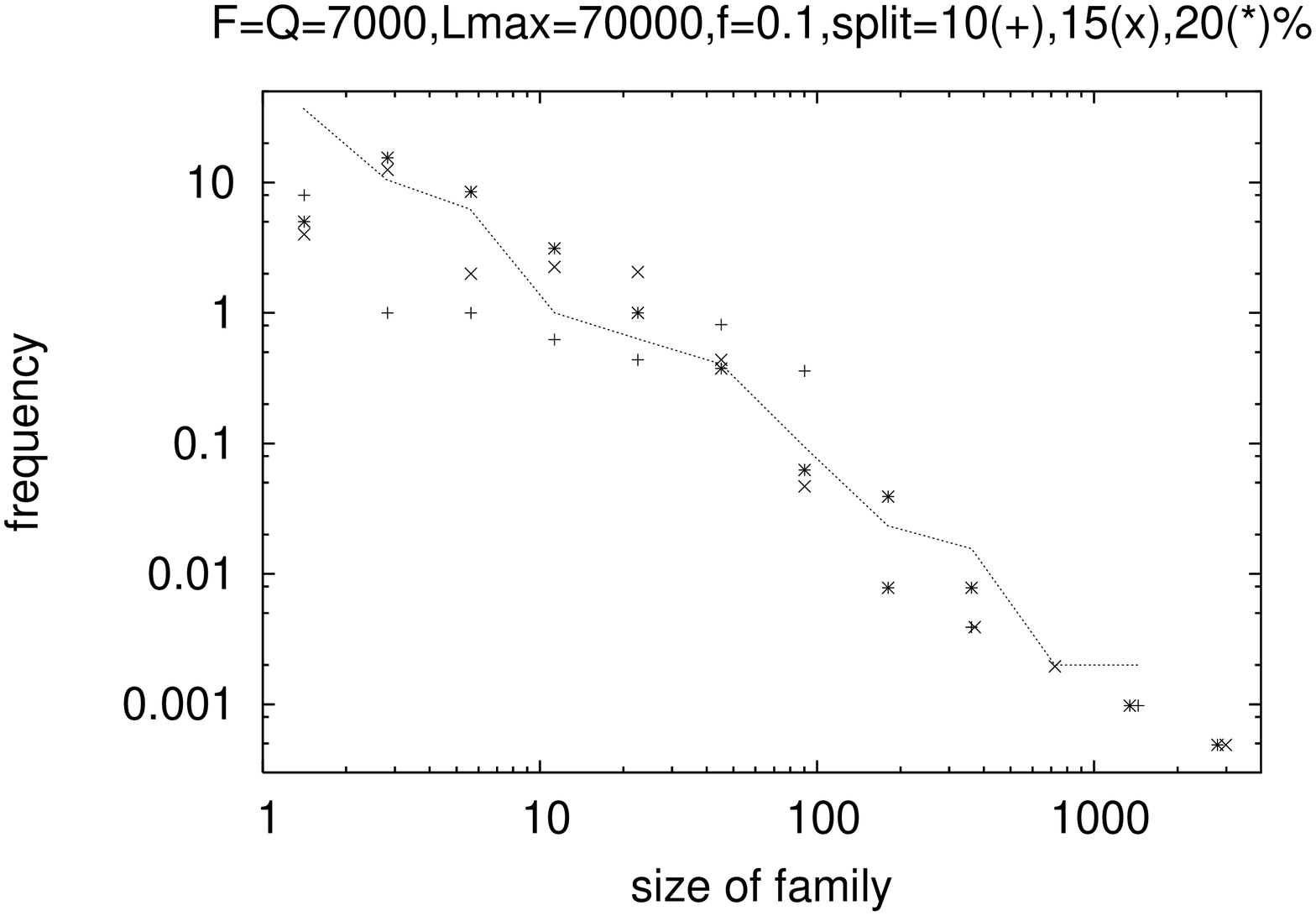}
\includegraphics[angle=-90,scale=0.3]{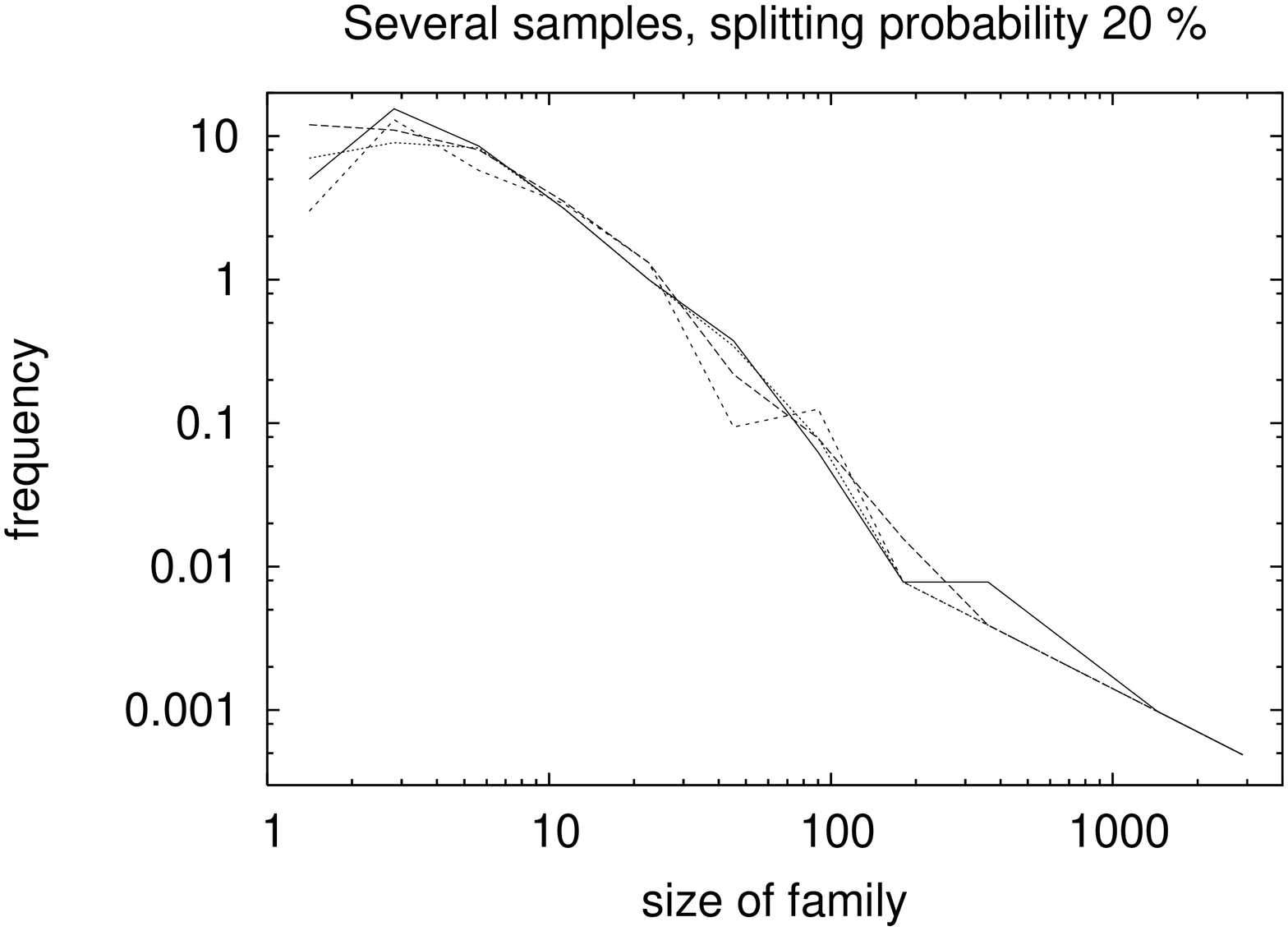}
\end{center}
\caption{Top: Three simulations with the modified definition of "family", the 
line corresponds to reality \cite{wichmann}. Bottom: Sample to sample 
fluctuations when only the random number seed is changed.
}
\end{figure}

\begin{figure}[hbt]
\begin{center}
\includegraphics[angle=-90,scale=0.5]{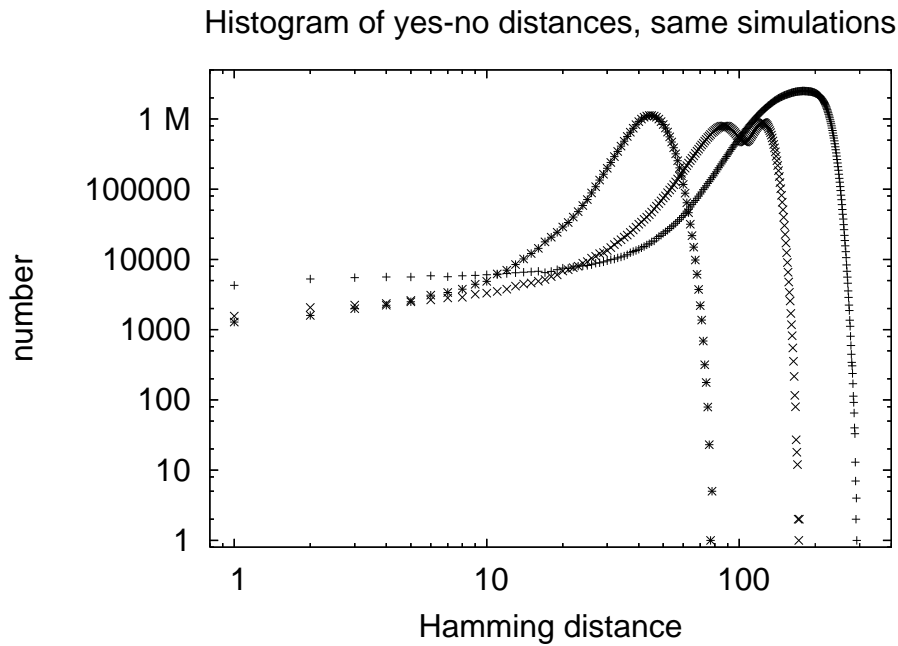}
\end{center}
\caption{Distribution of Hamming distances in the simulations of Fig.3 top.
See Fig.14 in \cite{tversky} for similar results.}
\end{figure}

\begin{figure}[hbt]
\begin{center}
\includegraphics[angle=-90,scale=0.5]{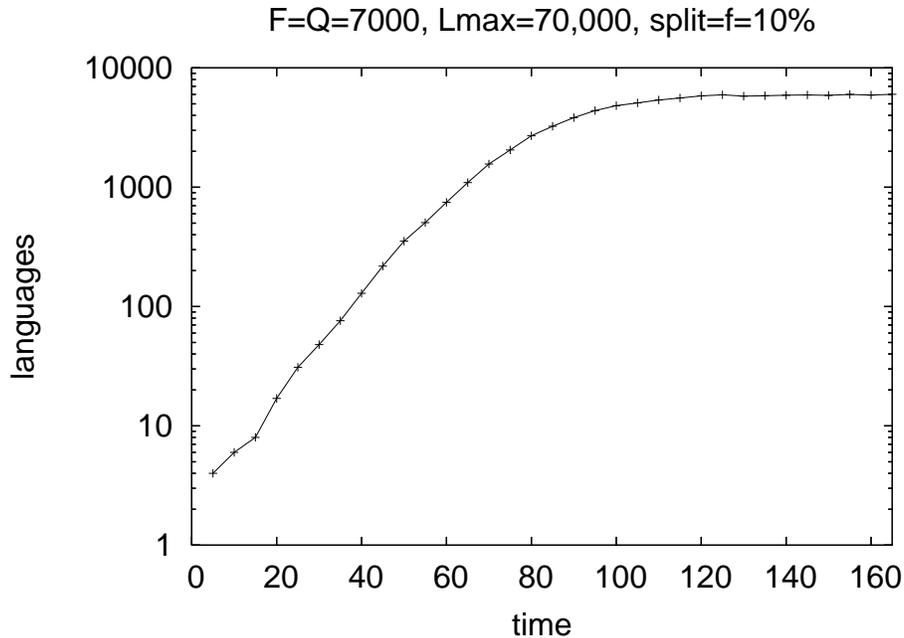}
\end{center}
\caption{Time variation of the number $L$ of languages in one of the samples 
of Fig.3 top.
}
\end{figure}

\begin{figure}[hbt]
\begin{center}
\includegraphics[angle=-90,scale=0.3]{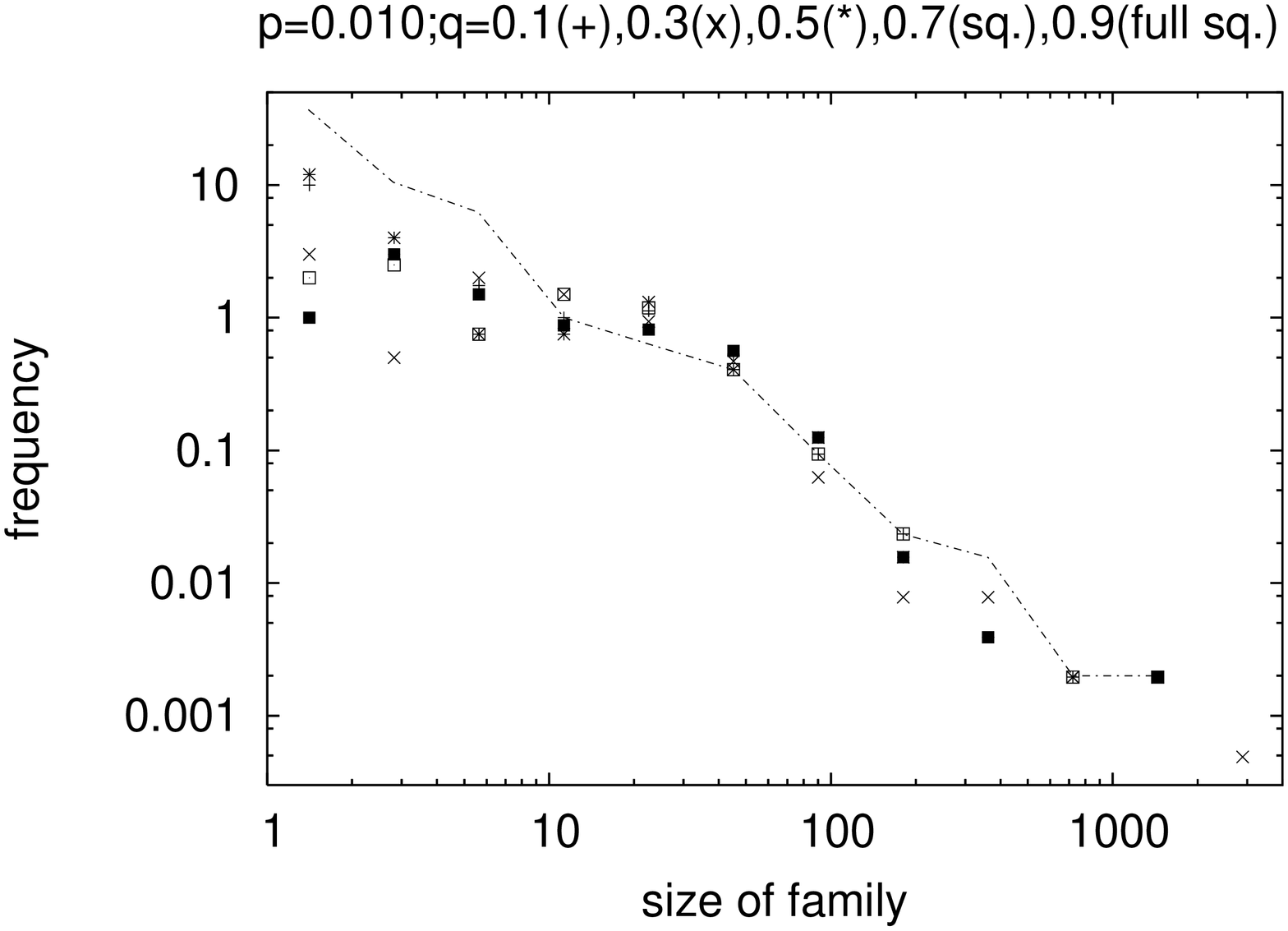}
\includegraphics[angle=-90,scale=0.3]{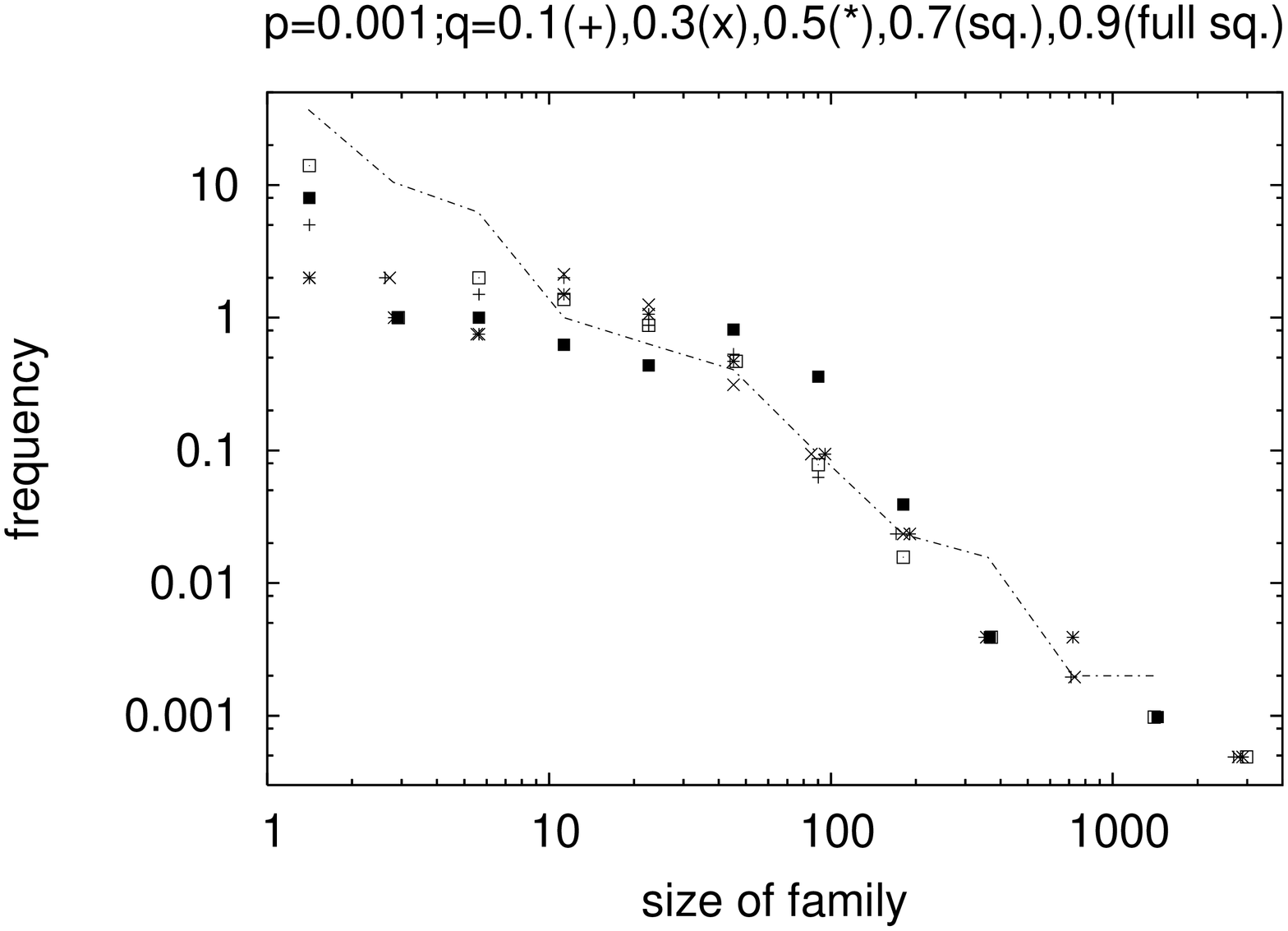}
\end{center}
\caption{Modified family size distributions for various diffusion probabilities
$q$, with $p = 0.01$ (top) and 0.001 (bottom); $F=Q=7000$. The lines again 
indicate reality \cite{wichmann}. 
}
\end{figure}

\begin{figure}[hbt]
\begin{center}
\includegraphics[angle=-90,scale=0.5]{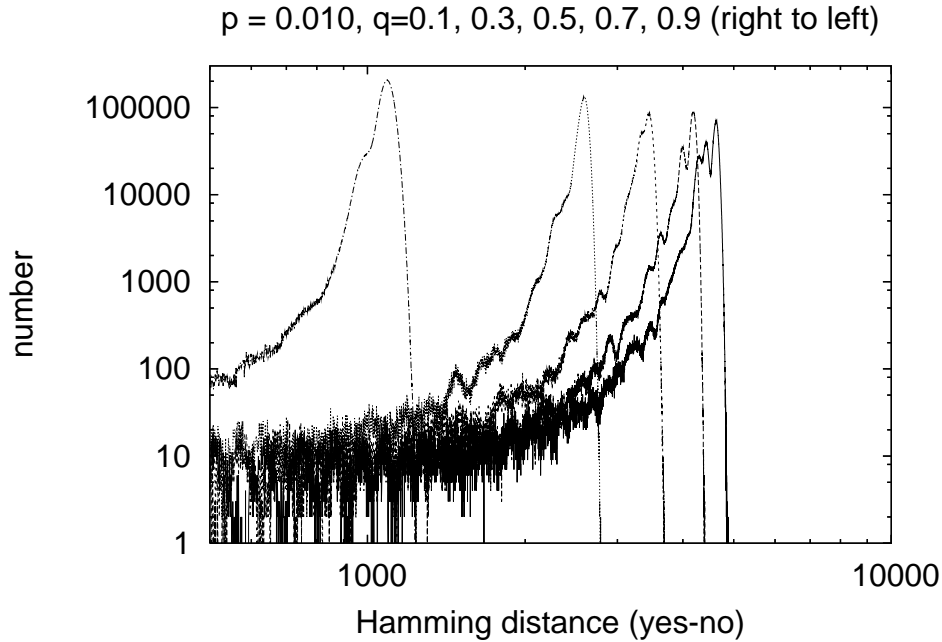}
\end{center}
\caption{Modified yes-no Hamming distance distributions for various diffusion 
probabilities $q$, with $p = 0.01, \; F=Q=7000$.
}
\end{figure}

\begin{figure}[hbt]!
\begin{center}
\includegraphics[angle=-90,scale=0.30]{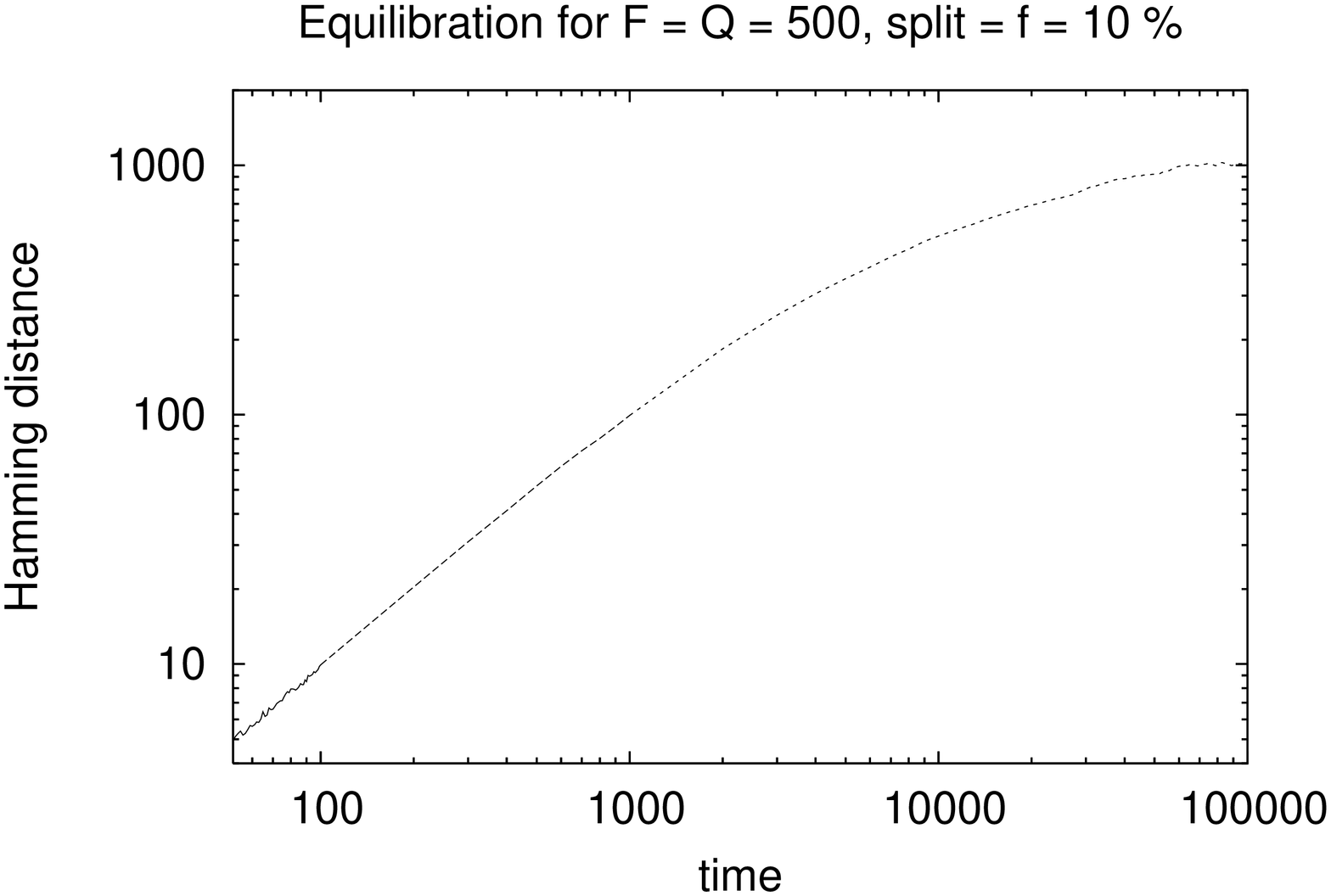}
\includegraphics[angle=-90,scale=0.30]{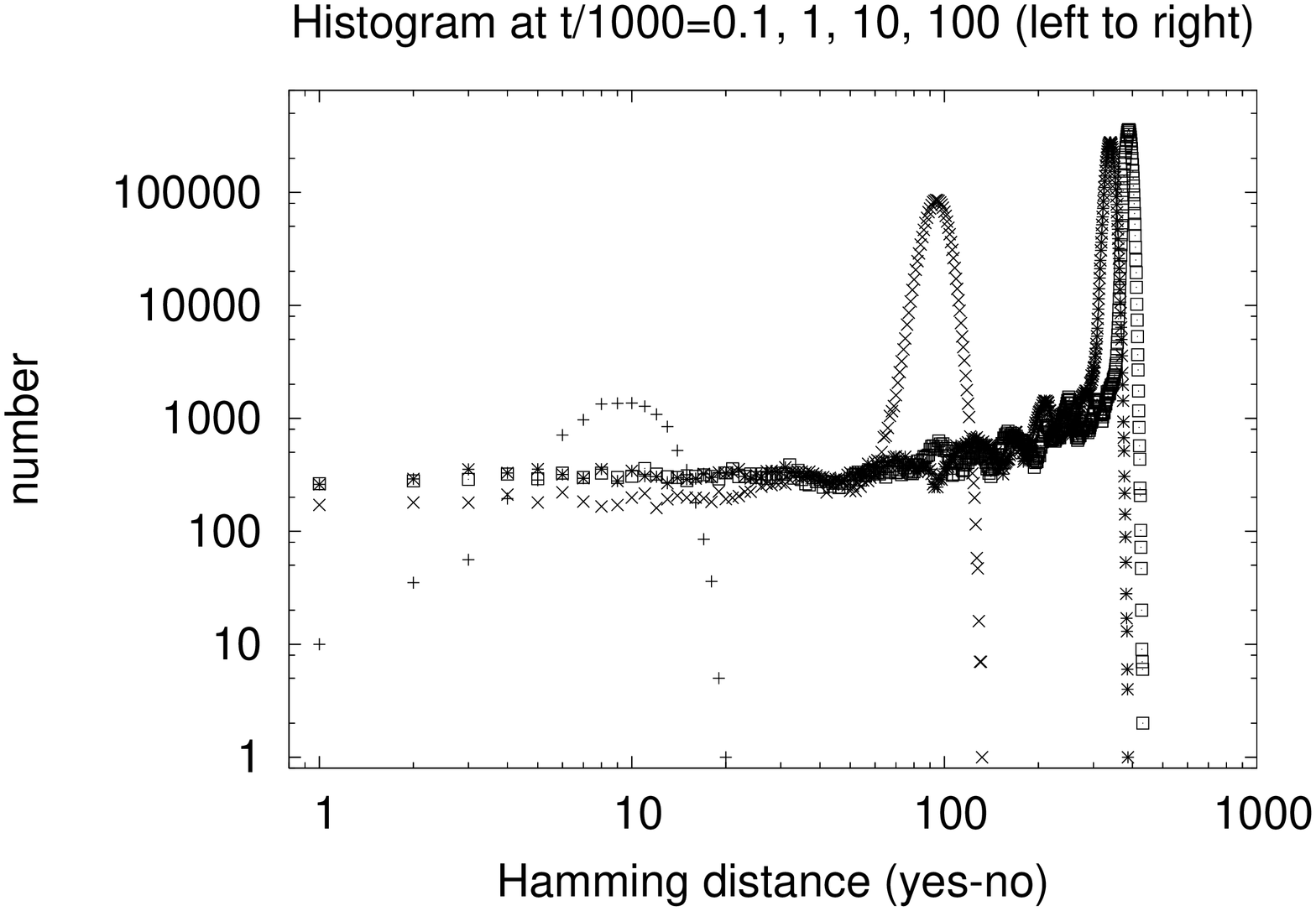}
\includegraphics[angle=-90,scale=0.30]{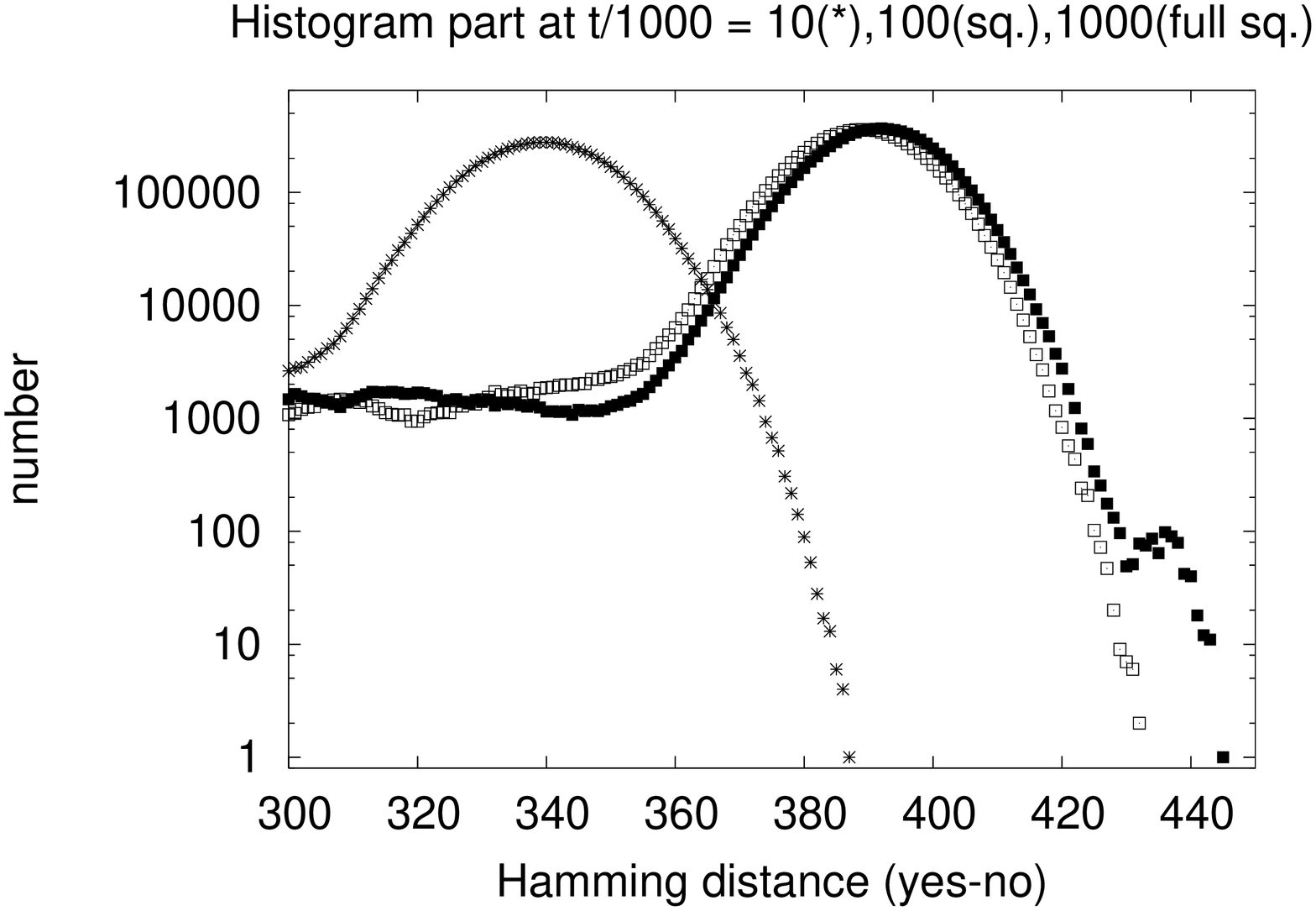}
\end{center}
\caption{Time variation in modified model of average Hamming (Manhattan) 
distance (top) and of Hamming distance distribution (centre and bottom) for 
small $F=Q=500$.
}
\end{figure}

\begin{figure}[hbt]
\begin{center}
\includegraphics[angle=-90,scale=0.5]{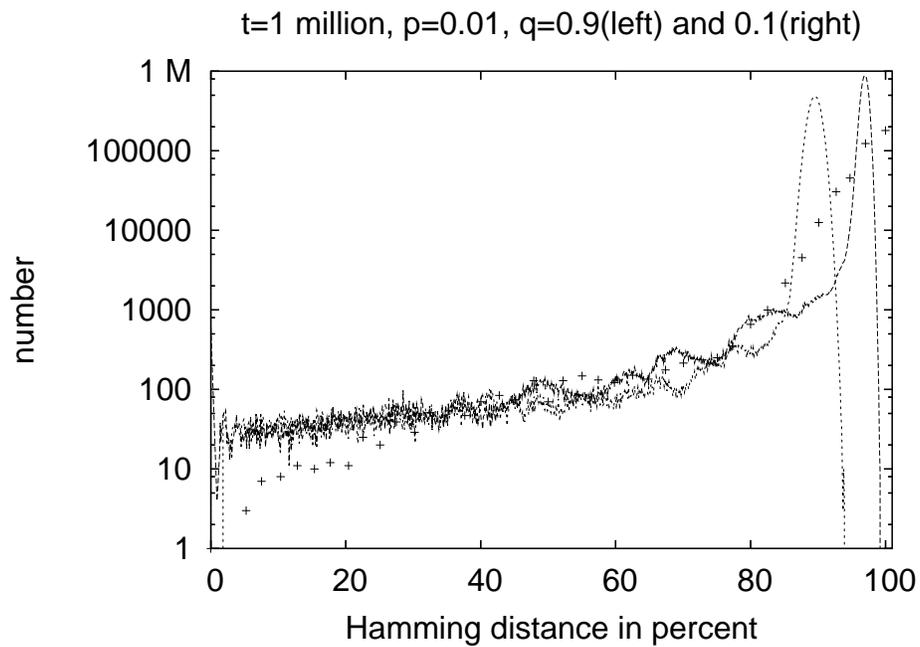}
\end{center}
\caption{Comparison of the reality (+) of automated lexicostatistics 
\cite{asjph,asjpb} with equilibrium simulations (yes-no; lines) for
$F=Q=1000$. Shorter simulations with only 100,000 and 300,000 iterations are not
significantly different. Thus $q \le 0.1$ roughly agrees with the 
lexicostatistical results.}
\end{figure}

\begin{figure}[hbt]
\begin{center}
\includegraphics[angle=-90,scale=0.41]{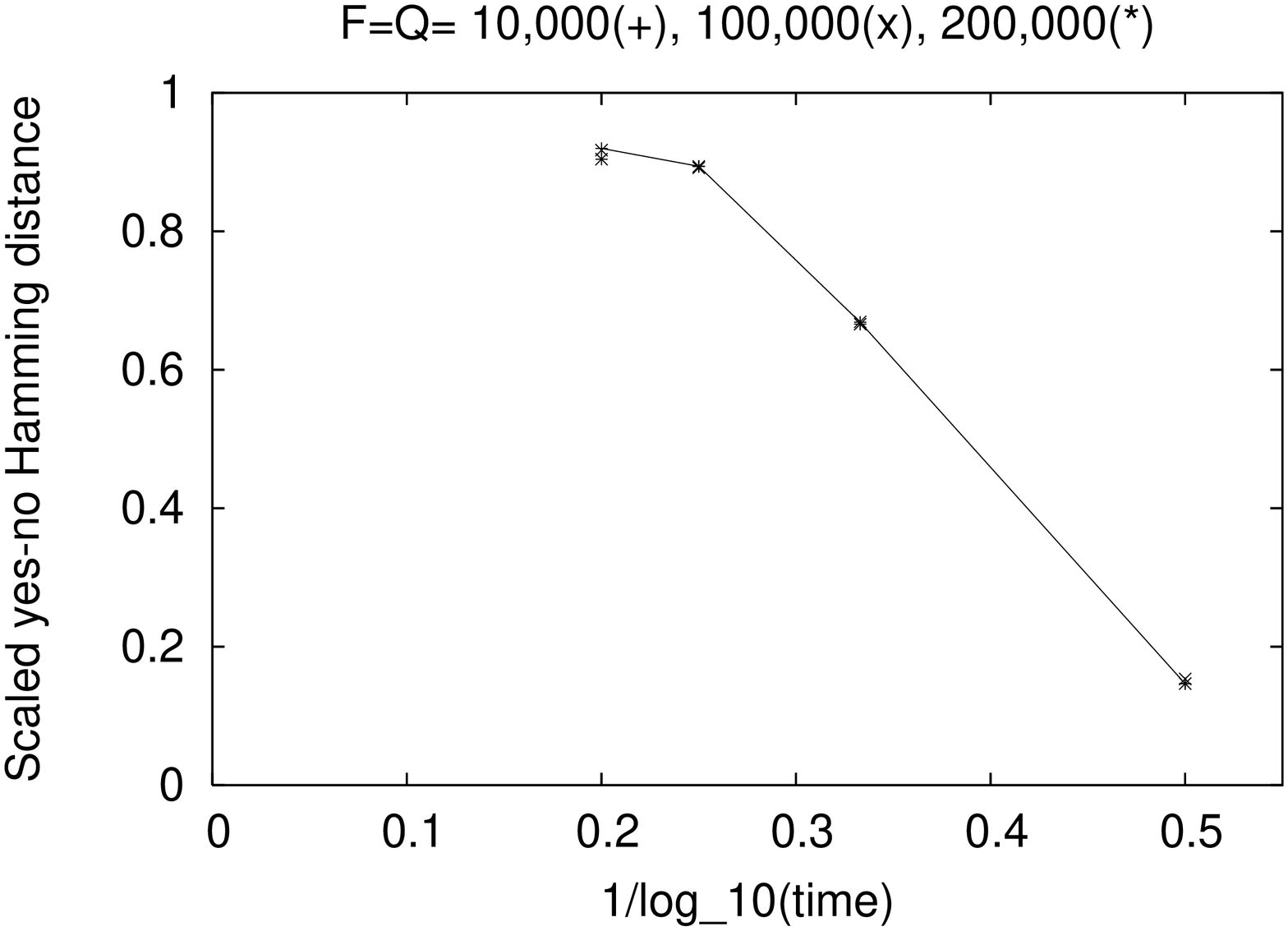}
\includegraphics[angle=-90,scale=0.41]{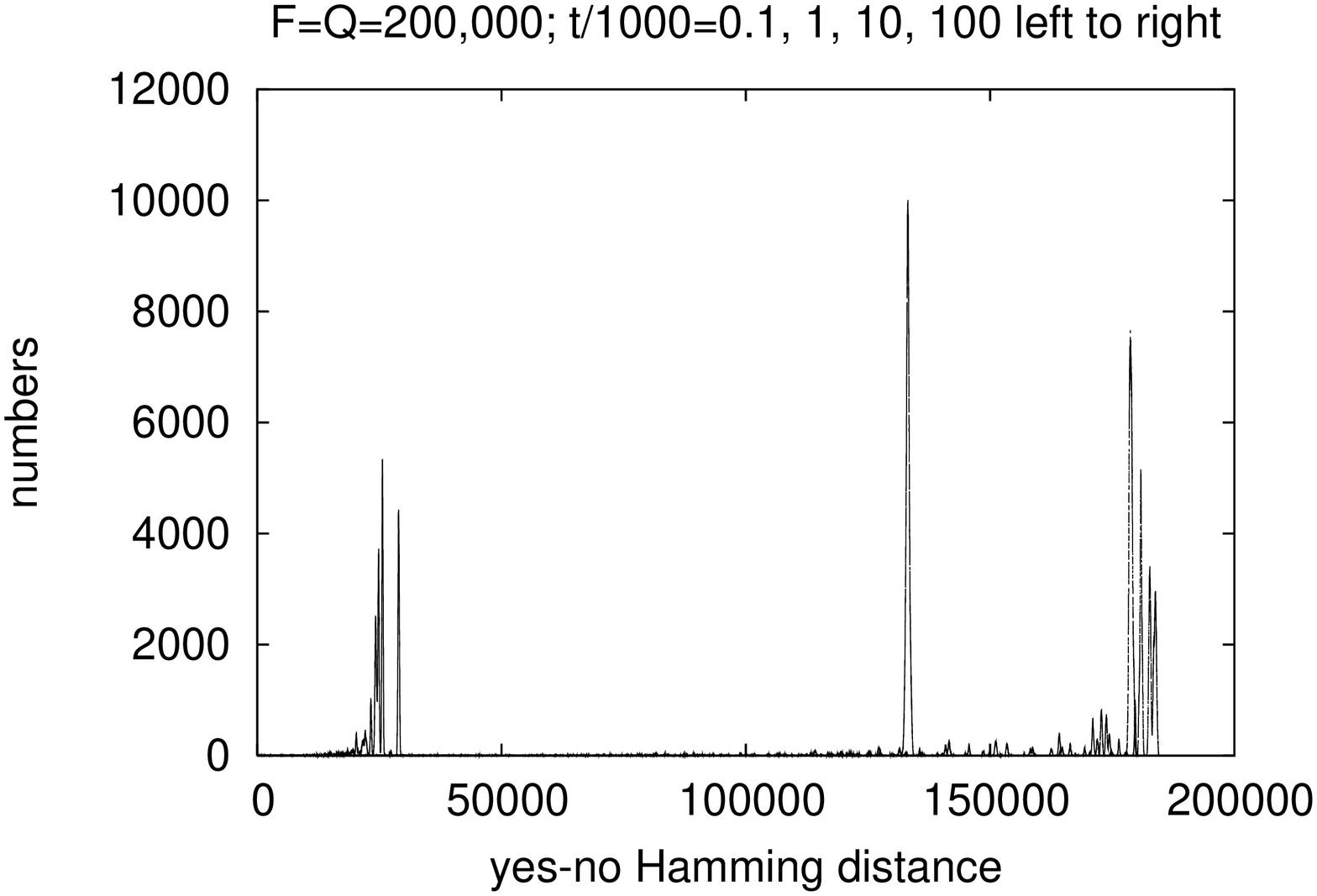}
\end{center}
\caption{
Top: Position of the maximum in the distribution of the yes-no Hamming distance,
scaled by the largest possible distance $F$, as a function of 1/log($t$), for
three large values of $F=Q$. Note the flattening for $t > 10^4$. The bottom 
part shows the whole distributions for the largest $F=Q=200,000$ (unscaled 
yes-now Hamming distance); sometimes there is a single peak, sometimes there
are several peaks close together.}

\end{figure}

In the present paper we regard grammatically
related words (e.g., life, live, lives, lived, living) as one "form",
and denote similarly related concepts by one "meaning". 
In the terminology of linguistics this corresponds to looking only at
lexical morphemes, ignoring various inflections and derivations.
Thus our $N$ languages
consist each of $F$ meanings and $Q$ forms; each meaning $i=1,2,\dots F$
is expressed by one form $S_i = 1,2,\dots Q$. One form may be
associated to several meanings, but no meaning is associated to
several forms. In reality the latter case, called homophony by linguists, does 
occur, but is somewhat rarer than the former case, termed polysemy.

The simulations allow for cases where a given meaning is not realised
in a given language, taking into account the sensitivity of  the lexical
inventories of languages to differences in cultural and natural
environments. Such an unrealized meaning could be denoted by $S_i=Q$.

We start with one language and one form, where all meanings have the central 
form $S_i = Q/2$; thus both the initial evolution of languages and their later
competition are simulated. Then we apply three processes: Change ("mutation") 
and diffusion ("transfer") of single features $S_i$ as in the Schulze
model \cite{ssw}, plus splitting \cite{tuncay} and merging of whole
languages. In this last (new?) process, two languages which agree in all
their $S_i$ at one time are regarded as one language from then on,
changing, diffusing, and splitting together, and potentially undergoing further merging with other languages. The real-world parallel to merging would be cases where incipient differences disappear shortly after they arise, something that happens when children change "wrong" forms popular among their peers to grown-up "correct" forms, when slang forms are invented and later forgotten again, when in-group varieties emerge and disappear, or when speakers of dialects shift to the standard variety. Different from the Schulze model 
and more similar to the Viviane und Tuncay models \cite{viviane,tuncay}, 
we no longer simulate each individual but only the language as a whole. 
Thus the "population" for one language no longer is part of this
model, and therefore, in contrast to the Schulze model, we have no shift from
languages spoken by few people to more widespread languages, only merging of 
similar variants, as mentioned above. And we
cannot determine a language size distribution, only a distribution for
the number of languages within one language family. Otherwise the new 
model is similar to the Schulze model. 

In the next section we define the parameters of this model, then
present our results, and in section 4 offer some modifications..

\section{Model}

A "language" is defined by a string of $F$ forms $S_i$, for each
meaning $i$ between 1 and $F$, where $S_i$ is an integer between 1 and
$Q$. Thus $Q^F$ different languages are possible.
At each iteration $t=1,2,\dots$ we go in the same order through all 
$N(t)$ different languages existing at that time $t$. Each of the $F$
language features at each iteration undergoes with probability $p$ a
change, which means with probability $q$ it takes over the
corresponding $S_i$ of a randomly selected language then existing in
the model, and with probability $1-q$ it changes its own $S_i$ by $\pm
1$ (but not below 1 or above $Q$). Also, at each iteration all language 
pairs agreeing in all their corresponding $S_i$ are merged into one 
language ($N \rightarrow N-1)$, and all languages surviving this
merging split with probability $s$ into two languages ($N \rightarrow
N+1)$ which from then on may diverge through change ($p$) and
diffusion ($q$). 

We start with $N=1$ language and during the first $\tau$ iterations
switch off the merging process (since otherwise we would always stay
at one, albeit changing, language). At the end of this time lag $t =
\tau$ each language is regarded as the founder of one language family,
comprising e.g. all Indo-European languages. The "size" of a family is
the number of different languages in it, arising from the later splitting
process.

In this sense the model combines language evolution and language
competition. A Fortran program is available upon request.

\section{Results} 

Mostly we worked with 7000 meanings and 7000 forms, at $p = 0.001, \; q = 0.9, 
\; s = 0.15,\; \tau = 40, \; 10^2$ iterations.
Figure 1 shows the size distribution for the families, which seems to be log-normal (parabola in 
this log-log plot). The plot is based on about 6000 languages in about 500 
language families which are realistic numbers \cite{wichmann} for the number of
languages in the present world. But since the distribution is 
more log-normal than power-law, the largest real families with about 1000 
languages are missing in this model. 

The Manhattan Hamming distance is the sum over all absolute differences between 
the corresponding $S_i$ of the two languages. 
Figure 2 gives the time dependence of the average Manhattan Hamming distances 
within one family and between different families; the second one is about twice 
as large as the first one, which seems reasonable \cite{holwich}.
All our simulations are non-equilibrium ones except for the number of families: 
The number of languages and the Hamming distances increase with time.

\section{Modifications}

A better result for the family size distribution, but a worse one for the 
Hamming distances, is obtained by a different family definition: whenever 
a new language is formed due to the splitting process, with probability $f=0.1$
it is regarded as the founder of a new language family, with all later
offspring languages belonging to it as long as they do not found a new family.
Then we see in Fig.3 a better straight line (power law), extending to larger 
family sizes and agreeing with reality \cite{wichmann}. The distributions 
of Hamming distances look as before, Fig.4, but now due to the random definition
of "family" there is no major difference between Hamming distances within one
and between different language families. In these simulations we adopted 
a Verhulst extinction probability $L/L_{\max}$ for each language at each 
iteration, where $L$ is the current number of languages and $L_{\max} = 70,000$.
As a result the number of languages may stabilise as in Fig.5, even though the
Hamming distances still increase; the number of families is about $100$.
Fig.6 shows that variations on the change and diffusion probabilities $p,q$ give
roughly the same family size distributions, and Fig.7 shows the variation 
of the Hamming distance histogram. (The numbers of families with only few
language is lower in the simulations of Figs. 4 and 7 than in reality; 
perhaps the simulations underestimate the extinctions of languages.)

The merging process was introduced to achieve, without our Verhulst extinction,
a stationary number of languages at long $t$ and very small $F$ and $Q$. 
For the large $F, Q$ and small $t$ used 
here this aim is not achievable. The family sizes without merging look 
similar (not shown). For small $F=Q=100,\ 200,\ 500$ we made up to a million
iterations and then see how the average Manhattan Hamming distance stabilises. 
Its yes-no
variant, counting only the number of different forms and not the amount of
their difference, gets close to its maximum value, Fig.8, for increasing time
(with extinction and merging).

In Fig. 9 we compare simulation results with empirical data from 859 languages collected as part of a project on automated lexicostatistics. For each language, a 40-word subset \cite{asjph} of the Swadesh 100-word basic vocabulary list \cite{swadesh} was transcribed in a standard orthography \cite{asjpb}. The distance between any two languages was defined as the percentage of attested words on the list that fail to match according to objective rules \cite{asjpb}. The data are well fit by the simulation with a low diffusion probability, $q = 0.1$: real and simulated languages differ typically in about 97\% of their words. (Only near
100 \% simulations and reality differ, since very few simulated languages
differ by 100 \%. With much smaller $F=Q=50$ this peak does not shift; with $q=0.01$ the peak shifts to 98 \%. Thus this discrepancy does not go away.) This result suggests that borrowing is not the source of most changes in basic vocabulary. 

 The greater variability of the data relative to the
 simulation may be attributed to two factors: first, random sampling 
 variability in the data, which are based on a 40-word sample from a 
 much larger true number of words; and second,
 the fact that likelihood of words
 matching in the data varies with the length of the words, while the simulation
 treats all words equally.

Finally, the largest languages $F=Q=7000$ used up to now are still smaller than 
the actual numbers of words used in normal speech (without technical expressions
and names). With a larger computer memory of 8 Gigabytes, and weeks 
of continuous simulation time, up to 200,000 meanings and forms were simulated
with $p=0.001, \; q=0.1, \; f=s=0.1, \; \tau=50$, 
surpassing the $10^5$ words learned by adulthood \cite{bloom}. Fig.10a shows
no significant changes in the scaled position of the maximum for the yes-no
Hamming distance, when $F=Q$ is increased (also for $F=Q=3000$ and 65,000; not 
shown). Fig.10b shows the whole distributions of the unscaled yes-no Hamming 
distance at $F=Q=200,000$ for various times $t$; at the right end of the plot 
we see a single maximum at $t=10^4$, followed to its right by three peaks of 
decreasing height at $t = 10^5$. Fig.10 thus confirms for larger $F=Q$ what was
seen already in Figs.8,9, that for long times $t$ a Hamming distance 
close to but below 100 percent is reached for the peak position.

\section{Summary}

Our model allowed for the first time for the simulation of thousands of meanings
and forms. The resulting family size distribution was (depending on definition)
close to log-normal or close to the power law of reality 
\cite{zanetteold,wichmann}, with large fluctuations.  The distribution of
Hamming distances had a skewed and realistic shape. With one of the definitions,
strong differences, as desired, were found between the Hamming distances of
languages within one family and between different families. 

\medskip
{\bf Acknowledgement}

\noindent
We thank a referee for drawing our attention to ref.\cite{bloom}.
For their efforts in gathering and analyzing the empirical data plotted in Fig. 
9, we would like to acknowledge the members of the Automatic Similarity Judgment
Project, in alphabetical order: Dik Bakker, Cecil
H. Brown, Pamela Brown, Hagen Jung, Robert Mailhammer, Andr\'e M\"uller, and
Viveka Velupillai. (Holman and Wichmann are also part of this project). A
list of languages used, a map of their distribution, references to data
sources, and two papers \cite{asjph,asjpb} can be accessed through the project 
web page:

\noindent
http://email.eva.mpg.de/$\sim$wichmann/ASJPHomePage.htm.

\end{document}